\documentclass[aps,prl,superscriptaddress,showpacs,preprintnumbers,amsmath,amssymb]{revtex4}
\usepackage{graphicx} 
\usepackage{dcolumn}  
\usepackage{xspace}
\usepackage{color}
\graphicspath{{ps}}

\newcommand{\bb}{\ensuremath{B \overline{B}}\xspace}

\newcommand{\btoetapi}{\ensuremath{B^+ \to \eta \pi^+}\xspace}
\newcommand{\btoetak}{\ensuremath{B^+ \to \eta K^+}\xspace}

\newcommand{\btoetakz}{\ensuremath{B^0 \to \eta K^0}\xspace}    
\newcommand{\btoetapiz}{\ensuremath{B^0 \to \eta \pi^{0}}\xspace}
\newcommand{\btoetaeta}{\ensuremath{B^0 \to \eta \eta}\xspace}

\newcommand{\etagg}{\ensuremath{\eta_{\gamma \gamma}}\xspace}
\newcommand{\etapi}{\ensuremath{\eta_{3\pi}}\xspace}

\newcommand{\de}{\ensuremath{\Delta E}\xspace}
\newcommand{\br}{\ensuremath{\mathcal{B}}\xspace}
\newcommand{\acp}{\ensuremath{\mathcal{A}_{CP}}\xspace}

\def\calL{{\mathcal L}}

\def\Mbc{M_{\rm bc}}
\def\LR{{\mathcal R}}

\begin{document}


\preprint{\vbox{ 
                 \hbox{Belle-Preprint 2004-38}
                 \hbox{KEK-Preprint 2004-79} 
}}

\title{ \quad\\[0.5cm] Measurements of Branching Fractions and $CP$ Asymmetries 
in $B\to \eta h$ Decays}

\affiliation{Budker Institute of Nuclear Physics, Novosibirsk}
\affiliation{Chiba University, Chiba}
\affiliation{Chonnam National University, Kwangju}
\affiliation{University of Cincinnati, Cincinnati, Ohio 45221}
\affiliation{Gyeongsang National University, Chinju}
\affiliation{University of Hawaii, Honolulu, Hawaii 96822}
\affiliation{High Energy Accelerator Research Organization (KEK), Tsukuba}
\affiliation{Hiroshima Institute of Technology, Hiroshima}
\affiliation{Institute of High Energy Physics, Vienna}
\affiliation{Institute for Theoretical and Experimental Physics, Moscow}
\affiliation{J. Stefan Institute, Ljubljana}
\affiliation{Kanagawa University, Yokohama}
\affiliation{Korea University, Seoul}
\affiliation{Kyungpook National University, Taegu}
\affiliation{Swiss Federal Institute of Technology of Lausanne, EPFL, Lausanne}
\affiliation{University of Ljubljana, Ljubljana}
\affiliation{University of Maribor, Maribor}
\affiliation{University of Melbourne, Victoria}
\affiliation{Nagoya University, Nagoya}
\affiliation{Nara Women's University, Nara}
\affiliation{National Central University, Chung-li}
\affiliation{National Kaohsiung Normal University, Kaohsiung}
\affiliation{National United University, Miao Li}
\affiliation{Department of Physics, National Taiwan University, Taipei}
\affiliation{H. Niewodniczanski Institute of Nuclear Physics, Krakow}
\affiliation{Nihon Dental College, Niigata}
\affiliation{Niigata University, Niigata}
\affiliation{Osaka City University, Osaka}
\affiliation{Osaka University, Osaka}
\affiliation{Panjab University, Chandigarh}
\affiliation{Peking University, Beijing}
\affiliation{Princeton University, Princeton, New Jersey 08545}
\affiliation{University of Science and Technology of China, Hefei}
\affiliation{Seoul National University, Seoul}
\affiliation{Sungkyunkwan University, Suwon}
\affiliation{University of Sydney, Sydney NSW}
\affiliation{Tata Institute of Fundamental Research, Bombay}
\affiliation{Toho University, Funabashi}
\affiliation{Tohoku Gakuin University, Tagajo}
\affiliation{Tohoku University, Sendai}
\affiliation{Department of Physics, University of Tokyo, Tokyo}
\affiliation{Tokyo Institute of Technology, Tokyo}
\affiliation{Tokyo Metropolitan University, Tokyo}
\affiliation{University of Tsukuba, Tsukuba}
\affiliation{Virginia Polytechnic Institute and State University, Blacksburg, Virginia 24061}
\affiliation{Yonsei University, Seoul}
 \author{P.~Chang}\affiliation{Department of Physics, National Taiwan University, Taipei} 
  \author{K.~Abe}\affiliation{High Energy Accelerator Research Organization (KEK), Tsukuba} 
  \author{K.~Abe}\affiliation{Tohoku Gakuin University, Tagajo} 
  \author{H.~Aihara}\affiliation{Department of Physics, University of Tokyo, Tokyo} 
  \author{M.~Akatsu}\affiliation{Nagoya University, Nagoya} 
  \author{Y.~Asano}\affiliation{University of Tsukuba, Tsukuba} 
  \author{T.~Aushev}\affiliation{Institute for Theoretical and Experimental Physics, Moscow} 
  \author{A.~M.~Bakich}\affiliation{University of Sydney, Sydney NSW} 
  \author{V.~Balagura}\affiliation{Institute for Theoretical and Experimental Physics, Moscow} 
  \author{Y.~Ban}\affiliation{Peking University, Beijing} 
% ...
\collaboration{The Belle Collaboration}

\begin{abstract}
We report measurements of $B$ to pseudoscalar-pseudoscalar decays with 
at least one $\eta$ meson in the final state using $140 \mathrm{~fb}^{-1}$ of 
data collected by the Belle detector at KEKB $e^+ e^-$ collider.  
We observe the decay  $\btoetapi$ and find  evidence of  $\btoetak$; the 
measured branching fractions are
$\br(\btoetapi) = ( 4.8^{+0.8}_{-0.7} \mathrm{(stat)} \pm 0.3
 \mathrm{(sys)}) \times 10^{-6}$ and
$\br(\btoetak) = ( 2.1\pm 0.6 \mathrm{(stat)} \pm 0.2
\mathrm{(sys)}) \times 10^{-6}$. Their corresponding $CP$ violating
asymmetries are measured to be $0.07\pm 0.15 \mathrm{(stat)} 
\pm 0.03 \mathrm{(sys)}$ for $\eta \pi^\pm$ and $-0.49\pm 0.31 \mathrm{(stat)} 
\pm 0.07  \mathrm{(sys)}$ for $\eta K^\pm$. No significant signals are found for
neutral $B\to \eta h$ decays. We report the following upper limits on branching 
fractions at the 90\% confidence level: 
$\br(\btoetakz)< 2.0\times 10^{-6}, \;
\br(\btoetapiz)<2.5 \times 10^{-6}$ and 
$\br (\btoetaeta) <2.0 \times 10^{-6}$.

\end{abstract}

\pacs{13.25.Hw, 12.15.Hh, 11.30.Er}

\maketitle

\tighten

{\renewcommand{\thefootnote}{\fnsymbol{footnote}}}
\setcounter{footnote}{0}
Charmless $B$ decays provide a  rich sample to understand  $B$ decay
dynamics and to search for $CP$ violation. An unexpectedly large 
$B\to\eta^\prime K$ branching fraction \cite{etapetac,etapk} 
has stimulated much theoretical
interest. It was suggested even before the $\eta^\prime K$ measurement that
 two $b\to s$ 
penguin amplitudes are constructive in  $B\to \eta^\prime K$ decays 
but destructive in $B\to \eta K$ \cite{lipkin}. The situation is reversed
for $B\to \eta^\prime K^*$ and $B\to \eta K^*$ decays. 
 Experimental results have more or 
less confirmed this picture; however, precise measurements of  branching
fractions are needed to quantitatively understand the contribution of each 
diagram. It was also pointed out that in the $\eta K$ mode the suppressed 
penguin amplitudes may interfere  with the CKM suppressed $b\to u$
(tree) amplitude and result in  direct $CP$ violation \cite{etakt}. 
The penguin-tree 
interference may also be large in $B^+\to\eta^\prime \pi^+$ \cite{CC} and 
$B^+\to\eta \pi^+$ decays; however, theoretical expectations for the 
partial rate asymmetry $(A_{CP})$ can be either positive or negative
\cite{etakt,etapit}. Recently, the BaBar Collaboration has reported  large 
negative $A_{CP}$ values in both $\eta K^+$ 
and $\eta \pi^+$, which are $\sim 2 \sigma$ away from zero \cite{babar}.
However, more data  are needed to verify these large $CP$ violating 
asymmetries. 
Furthermore, branching fractions and partial rate asymmetries in charmless 
$B$ decays can be used to understand the tree and penguin contributions and
provide constraints on the third unitarity triangle angle $\phi_3$ 
\cite{rosner}.      

In this paper, we report  measurements of branching fractions and partial
rate asymmetries for  $B\to\eta h$ decays, where $h$ could be a 
$K,\; \pi$ or $\eta$ meson. The partial rate asymmetry is measured for the
charged $B$ decays and  defined to be:
\begin{eqnarray*}
\acp=\frac{N( B^- \to \eta h^-)-N(B^+ \to \eta h^+)}
{N(B^- \to \eta h^-)+N(B^+ \to \eta h^+)},
\end{eqnarray*}
where $N(B^-)$ is the yield for the $B^- \to \eta h^-$ decay and
$N(B^+)$ denotes that of the charge conjugate mode.
 The data sample consists of 152 
million \bb pairs (140 fb$^{-1}$) collected
with the Belle detector at the KEKB $e^+e^-$ asymmetric-energy  
(3.5 on 8~GeV) collider~\cite{KEKB} operating at the $\Upsilon(4S)$ resonance.

The Belle detector is a large-solid-angle magnetic
spectrometer that
consists of a three-layer silicon vertex detector (SVD),
a 50-layer central drift chamber (CDC), an array of
aerogel threshold \v{C}erenkov counters (ACC),
a barrel-like arrangement of time-of-flight
scintillation counters (TOF), and an electromagnetic calorimeter (ECL)
comprised of CsI(Tl) crystals located inside
a superconducting solenoid coil that provides a 1.5~T
magnetic field.  An iron flux-return located outside of
the coil is instrumented to detect $K_L^0$ mesons and to identify
muons (KLM).  The detector
is described in detail elsewhere~\cite{Belle}.

Two $\eta$ decay channels are considered in this analysis: 
$\eta\to \gamma\gamma$ ($\eta_{\gamma\gamma}$) and $\eta\to \pi^+\pi^-\pi^0$
 ($\etapi$).  In the $\etagg$ reconstruction, each photon is required to have 
a minimum laboratory energy of 50 MeV and the energy asymmetry, 
defined as the absolute value of the energy difference in the laboratory frame
between the two photons divided by their energy sum, 
must be less than 0.9. 
Furthermore, we remove  $\eta$ candidates if either one of the daughter 
photons can pair with any other photon  to form a $\pi^0$ candidate. Candidate $\etapi$ mesons are reconstructed by combining a $\pi^0$ with a pair
of oppositely charged tracks that originate from the interaction point (IP).
We make the following requirements for the invariant mass  on 
the $\eta$ candidates: 516 MeV/$c^2 < M_{\gamma\gamma} < 569$ MeV/$c^2$ for 
$\etagg$ and 539 MeV/$c^2 < M_{3\pi} <556$ MeV/$c^2$ for $\etapi$. After the 
selection of each 
candidate, an $\eta$ mass constraint is implemented  by readjusting the momenta
of the daughter particles.   

Candidate neutral pions are selected by requiring the
two-photon invariant mass to be in the mass window between 115 MeV/$c^2$ and 
152 MeV/$c^2$. The momentum of each photon is then readjusted to
constrain the mass of the photon pair to be the nominal $\pi^0$ mass. 
To reduce the low energy photon background,  
each photon is required to have a minimum energy of 50 MeV and the 
$\pi^0$ momentum 
must be above 250 MeV/$c$ in the laboratory frame. 
Charged tracks are required to come from the IP.  Charged kaons and pions 
that form $B$ candidates with $\eta$ mesons
are identified by combining information from the CDC ($dE/dx$),
the TOF and the ACC to form a $K(\pi)$ likelihood $L_K(L_\pi)$. 
Discrimination between kaons and pions is achieved through
the likelihood ratio $L_{K}$/($L_{\pi}+L_{K}$). Charged tracks with  
likelihood ratios greater than 0.6 are regarded as kaons, and less than 0.4
as pions. The typical kaon and pion identification efficiencies for 2.5 GeV/$c$
tracks are  $(85.0\pm 0.2)$\% and $(89.3\pm 0.2)$\%, respectively. With the
same track momentum, the rate for  pions to be misidentified as kaons is 
$(7.3 \pm 0.2)$\% while the rate for  kaons  to be misidentified as pions is
$(10.6 \pm 0.2)$\%.
Furthermore, charged tracks that are positively identified as 
electrons or muons are rejected. $K^0_S$ candidates 
 are reconstructed from  pairs of oppositely charged tracks
 with invariant mass ($M_{\pi\pi}$) between 480 to 516 MeV/$c^2$.
Each candidate must have a displaced vertex with a flight direction
 consistent with a $K^0_S$ originating from the IP.

Candidate $B$ mesons are identified using the beam constrained mass,
$\Mbc =  \sqrt{E^2_{\mbox{\scriptsize beam}} - P_B^2}$,
and the energy difference, $\Delta E = E_B  - E_{\mbox{\scriptsize beam}}$,
where $E_{\mbox{\scriptsize beam}}$ is the run-dependent beam energy in the
$\Upsilon(4S)$ rest frame and  is determined  from
$B\to D^{(*)}\pi$ events,
and $P_{B}$ and $E_B$ are the momentum and energy of  the
$B$ candidate in the $\Upsilon(4S)$ rest frame. The resolutions on $\Mbc$ and 
$\de$ are around 3 MeV/$c^2$ and $\sim$ 20--30 MeV, respectively.  
Events with $\Mbc >5.2$ GeV/$c^2$ and $|\de|<0.3$ GeV are selected for the 
 analysis.   

The dominant background comes from the $e^+e^-\rightarrow q\bar{q}$ continuum, 
where $q= u, d, s$ or $c$. To distinguish signal from the jet-like continuum 
background,  event shape variables and the $B$ flavor tagging information 
are employed. We form a Fisher discriminant \cite{fisher} from seven
variables that quantify event topology.  The Fisher variables include
the angle $\theta_T$ between the thrust axis \cite{thrust} of the $B$ candidate
and the thrust axis of the rest of the event, five modified Fox-Wolfram
 moments \cite{sfw}, and a measure of the momentum transverse to the event 
thrust axis ($S_\perp$)  \cite{sperp}. The  probability density functions (PDF)
for this discriminant  and $\cos\theta_B$, where $\theta_B$ is the angle
between the $B$ flight direction and the beam direction
in the $\Upsilon(4S)$ rest frame,  are obtained using events in  signal 
Monte Carlo (MC) and data with $\Mbc< 5.26$ GeV/$c^2$ for signal and 
$q \bar{q}$ background, respectively. These two variables are then combined to 
form a likelihood ratio $\LR = {\calL}_s/({\calL}_s + {\calL}_{q \bar{q}})$,
where ${\calL}_{s (q \bar{q})}$ is the product of signal ($q \bar{q}$) 
probability densities.

Additional background discrimination is provided by the quality of 
the$B$ flavor tagging of the accompanying $B$ meson.
We use the standard Belle $B$ tagging package \cite{tagging},  
which gives two outputs:
a discrete variable $(q)$ indicating the $B$ flavor and a dilution factor ($r$)
ranging from zero for no flavor information to unity for unambiguous
flavor assignment. We divide the data into six $r$ regions.
Continuum suppression is achieved by applying a mode dependent requirement on 
$\LR$ for events in each $r$ region based on 
$N_s^{\rm exp}/\sqrt{N_s^{\rm exp}+N_{q\bar{q}}^{\rm exp}}$,
where $N_s^{\rm exp}$ is the expected signal  from MC and
$N_{q\bar{q}}^{\rm exp}$ denotes the number of background events estimated from data.
This $\LR$ requirement retains 58--86\% of the signal  while reducing 
96--82\% of 
the background. From MC all other backgrounds are found to be negligible 
 except for the 
$\eta K^+ \leftrightarrow \eta \pi^+$ reflection, due to 
$K^+\leftrightarrow \pi^+$ 
misidentification, and the $\eta K^*(892) (\eta\rho(770))$ feed-down to the
$\eta K (\eta \pi)$ modes. We include these two components in the fit used 
to extract the signal.  

The signal yields and branching fractions are obtained using an extended 
unbinned maximum-likelihood (ML) fit with input variables $\Mbc$ and $\de$.
The likelihood is defined as: 
\begin{eqnarray*}
  \mathcal{L} = {\rm exp}\; (-\sum_j N_j) \prod^N_i \; [\; \sum N_j P_j(\Mbc{}_i,\de{}_i) \; ],
\end{eqnarray*}
where $N_j$ is the yield of category $j$ (signal, continuum background, 
reflection, $\eta K^*/\eta\rho$), $P_j (\Mbc{}_i,\de{}_i)$ is the 
probability density for the $i$th event and $N$ is the total number of events.  
The PDFs for the signal, the reflection background and the $\eta K^*/\eta \rho$
feed-down are modeled with two-dimensional $\Mbc$-$\de$ smooth functions obtained
using MC. The signal peak positions and resolutions in $\Mbc$ and $\de$
are adjusted according to the data-MC differences using large control samples 
of $B\to D\pi$  and $\overline{D}{}^0\to K^+\pi^-\pi^0/\pi^0\pi^0$ decays. 
The continuum background in $\de$ is described by a first or second order
polynomial while the $\Mbc$ distribution is parameterized by an 
ARGUS function, $f(x) = x \sqrt{1-x^2}\;{\rm exp}\;[ -\xi (1-x^2)]$, where 
$x$ is $\Mbc$ divided by half of the total center of mass energy \cite{argus}. 
Thus the continuum 
PDF is the product of an ARGUS function and a polynomial, where
$\xi$ and the coefficients of the polynomial are free parameters. 
Since $B\to\eta K^*$ branching fractions are well measured 
($\sim 20 \times 10^{-6}$) \cite{etapetac,etakst}, their feed-down  
to the $\eta K$ modes are fixed from MC in the likelihood fit. Since  the 
decay $B^+\to \eta \rho^+$  is experimentally poorly constrained, the amount of this background  in the
$\eta \pi$ modes is allowed to float in the fit. In the charged $B$ modes,
the normalizations of the reflections are fixed to  expectations 
based on the $B^+\to \eta K^+$ and $B^+\to \eta \pi^+$  branching fractions and 
$K^+\leftrightarrow \pi^+$ fake rates, measured using   
$\overline{D}{}^0 \to K^+\pi^-$ data. The reflection yield is first estimated
with the assumed $\eta K^+$ and $\eta\pi^+$ branching fractions and is then
recalculated according to our measured branching fractions. No $\bb$ 
contributions
are included for the $B^0\to \eta \eta$ mode.

\begin{table*}[th]
\caption{ Detection efficiency ($\epsilon$), product of daughter branching 
fractions ( $\prod \br_i$), yield,
significance (Sig.), measured branching fraction ($\mathcal{B}$), 
the 90\% C.L. upper limit (UL) and $A_{CP}$ for the $B\to \eta h$ decays. 
The first errors
in columns 4, 6 and 8 are statistical and the second errors are systematic.}

\begin{tabular}{lccccccc} \hline\hline
Mode & $\epsilon(\%)$ & $\prod \br_i (\%)$ &  Yield & \hspace{0.2cm}Sig. 
\hspace{0.2cm} & $\mathcal{B} (10^{-6})$  & UL$(10^{-6}$)& $A_{CP}$ \\
\hline
\hline
 \btoetapi &
        & & & 8.0 & $4.8\pm 0.7\pm 0.3$ &
         & $0.07\pm 0.15 \pm 0.03$
        \\
\hspace{0.4cm}$\etagg \pi^+$ &
        $23.3$ &39.4 & $73.4^{+13.5}_{-12.7}\pm 2.0$ & 7.0 & $5.3^{+1.0}_{-0.9}
\pm 0.3$  &  & $0.11 \pm 0.17 \pm 0.03$
        \\
\hspace{0.4cm}$\eta_{3\pi} \pi^+$ &
        $14.8$ & 22.6& $19.6^{+7.0}_{-6.1}\pm 0.7$ & 3.9 & $3.8^{+1.4}_{-1.2} \pm 0.3$
  &          & $-0.11^{+0.35+0.04}_{-0.33-0.05}$
        \\
 \btoetak &
      & & & 3.7 & $2.1\pm 0.6 \pm 0.2$ &  & $-0.49\pm 0.31 \pm 0.07$ \\
  \hspace{0.4cm}$\etagg K^+$ &
       $21.1$ & $39.4$ & $28.0^{+10.0}_{-9.1}\pm 1.6$ & $3.3$ & $2.2^{+0.8}_{-0.7}\pm
       0.2$ &  & $-0.45^{+0.35}_{-0.31}\pm 0.07$ \\
  \hspace{0.4cm}$\eta_{3\pi} K^+$ &
       $13.8$ & $22.6$ & $7.4^{+5.4}_{-4.5}\pm 0.5$ & 1.7 & $1.5^{+1.1}_{-0.9}\pm 0.2$
       &  & $-0.78^{+1.03+0.11}_{-0.76-0.12}$ \\
 $B^0\to \eta K^0$ &
      & & & 0.4 & $0.3^{+0.9}_{-0.7}\pm 0.1$ & $<2.0$ & \\
  \hspace{0.4cm}$\etagg K^0$ & $22.9$ & $13.6$ & $-1.9^{+4.3}_{-3.1}\pm 0.3$ 
      &$-$  & $-0.4^{+0.9}_{-0.7}\pm 0.1$ & & \\
  \hspace{0.4cm}$\eta_{3\pi} K^0$ & $12.2$ & $7.8$ & $3.5^{+3.6}_{-2.7}\pm 0.2$
      &  1.3& $2.4^{+2.5}_{-1.9}\pm \pm 0.3$ & &\\
 $B^0\to \eta \pi^0$ &
      & & & 1.8 &  $1.2\pm 0.7\pm 0.1 $ & $<2.5$ & \\
  \hspace{0.4cm}$\etagg \pi^0$ & 17.0 & 39.0 & 
       $18.2^{+8.9+0.8}_{-8.0-0.7}$ & 2.4 & $1.8^{+0.9}_{-0.8}\pm 0.2$ & & \\
  \hspace{0.4cm}$\eta_{3\pi} \pi^0$ &11.2 & 22.3 & $-3.0^{+5.0}_{-4.0}\pm 0.3$ 
      &$-$ & $-0.8^{+1.3}_{-0.8}\pm 0.1$& & \\
 $B^0\to \eta \eta$ &
     & & &1.1 & $0.7^{+0.7}_{-0.6}\pm 0.1$  &$<2.0$ &  \\
  \hspace{0.4cm}$\etagg \etagg$ & 16.9 & 15.5 & $-1.5^{+2.7}_{-1.6}\pm 0.1$ & 
         $-$   & $-0.4^{+0.7}_{-0.4}\pm 0.0$ & & \\
  \hspace{0.4cm}$\etagg \eta_{3\pi}$ & 11.3 & 17.8 &$7.3^{+4.5}_{-4.0}\pm 0.2$ 
         &2.2 & $2.3^{+1.4}_{-1.2}\pm 0.2$ &  & \\
  \hspace{0.4cm}$\eta_{3\pi}\eta_{3\pi}$ & 7.7 &5.1&$0.3^{+2.0}_{-1.2}\pm 0.1$ &
         0.2 & $0.5^{+3.1}_{-1.9}\pm 0.1$ & & \\
\hline
\hline
\end{tabular}
\label{tab:result}
\end{table*}

In Table ~\ref{tab:result}  we show the measured branching
 fractions for each decay mode
as well as  other quantities associated with the measurements.
The efficiency for each mode is determined using  MC simulation and
corrected for the discrepancy between data and MC using the control samples. 
The only discrepancy we find is the performance of particle identification, 
which
results in a 4.3\% correction for the  $\eta \pi^+$ mode and 1.7\% for 
$B\to \eta K^+$.  
The combined branching fraction of the two $\eta$ decay modes is obtained 
from a simultaneous  likelihood fit to all the sub-samples with a common 
branching fraction. 
Systematic uncertainties in the fit due to the uncertainties in the signal PDFs
are estimated by performing the fit after varying their peak positions and 
resolutions by one standard deviation. In the $\eta K$ modes, we also vary the expected $\eta K^*$ 
feed-down by one standard deviation to check the yield difference. 
The quadratic 
sum of the deviations from the central value gives the systematic 
uncertainty in the fit,
which ranges from 3\% to 6\%. For each systematic check, the 
statistical significance is taken as the square root of
the difference between the value of $-2\ln\calL$ for zero signal yield and the
best-fit value. We regard the smallest value as our significance incuding 
the systematic uncertainty. 
The number of $B^+B^-$ and $B^0\overline{B}{}^0$ pairs are assumed to be
equal.

The performance of the $\LR$ requirement is 
studied by 
checking the data-MC efficiency ratio using the $B^+\to \overline{D}{}^0 \pi^+$ 
control sample. The obtained error is 2.4--3.5\%. The
 systematic errors on the charged track reconstruction
are estimated to be around $1$\% per track using  partially
reconstructed $D^*$ events, and  verified by comparing the ratio of
$\eta\to \pi^+\pi^-\pi^0$ to $\eta\to \gamma\gamma$
in data with MC expectations. The $\pi^0$ and $\eta_{\gamma\gamma}$ 
reconstruction efficiency is verified by comparing the $\pi^0$
decay angular distribution with the MC prediction, and by measuring the ratio 
of the branching fractions for the two $\eta$ decay channels: 
$\eta\to \gamma\gamma$
and $\eta\to \pi^0\pi^0\pi^0$. We assign 3.5\% error for the 
$\pi^0$ and $\etagg$ reconstruction. The $K_S^0$ reconstruction is verified by 
comparing the  ratio of $D^+\to K_S^0\pi^+$ and $D^+\to K^-\pi^+\pi^+$ yields. 
The resulting $K_S^0$ detection systematic error is 
4.4\%. The 
uncertainty in the number of $\bb$ events is 1\%. The final systematic error is
obtained by first summing all correlated errors linearly and then quadratically
summing the uncorrelated errors.

\begin{figure}[htb]
\includegraphics[width=0.7\textwidth]{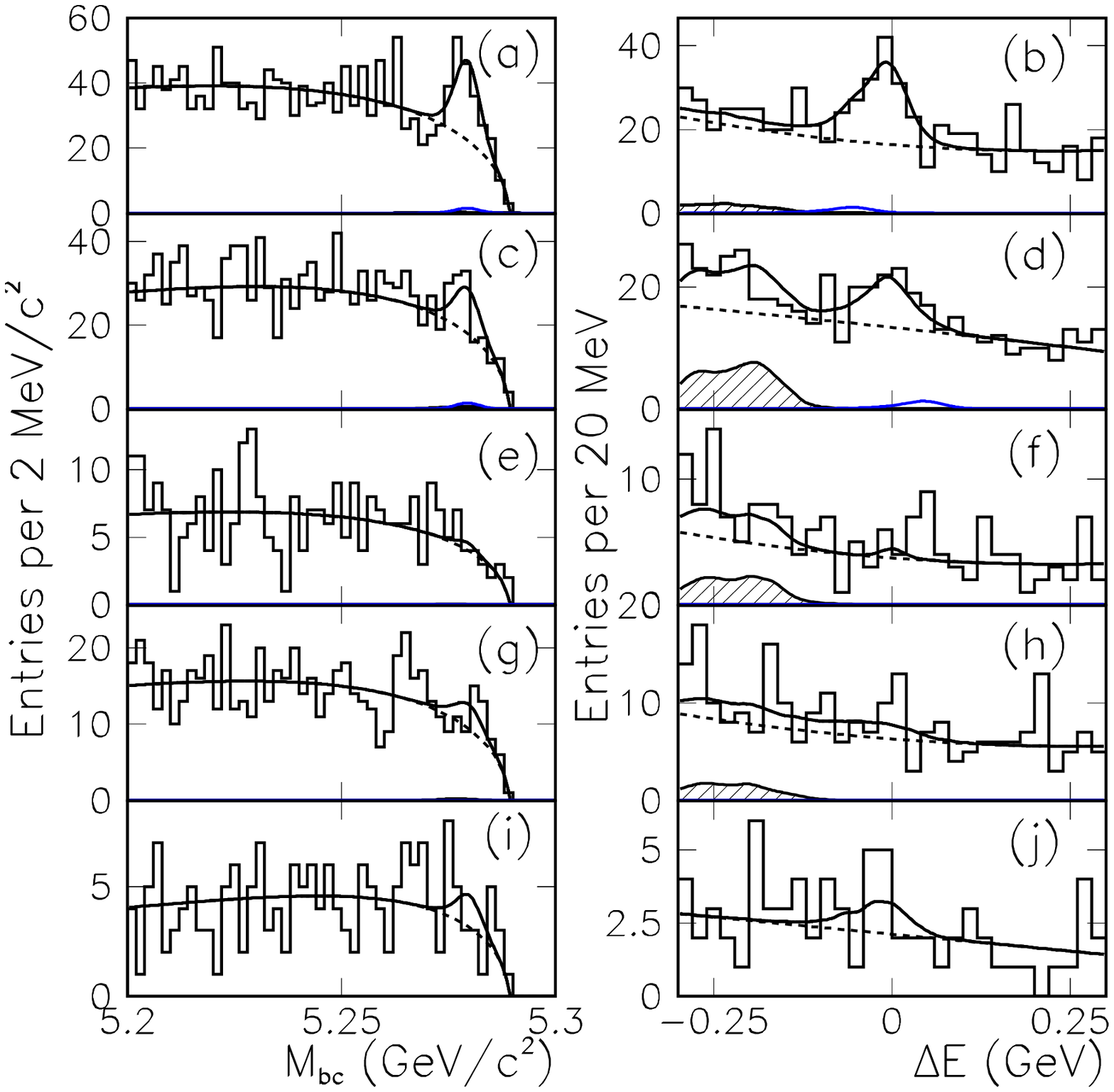}
\caption{$\Mbc$ and $\de$ projections for (a,b) $\btoetapi$, (c,d) $\btoetak$,
(e,f) $\btoetakz$, (g,h) $\btoetapiz$ and (i,j) $\btoetaeta$ decays with 
the $\etagg$ and $\etapi$ modes combined. Open
histograms are data, solid curves are the fit functions, dashed lines show
the continuum contributions and shaded histograms are the $\eta K^*/\eta \rho$
contributions.  The small contributions around $\Mbc = 5.28$ GeV/$c^2$ and 
$\de =\pm 0.05$
GeV in (a)-(d) are the backgrounds from $\btoetak$ and $\btoetapi$.  
}
\label{fig:mbde}
\end{figure}

 Figure \ref{fig:mbde} shows the $\Mbc$ and $\de$ projections after requiring
events to satisfy $-0.10$ GeV $<\de<0.08$ GeV ($-0.15$ GeV $<\de<0.10$ GeV for 
the $\etagg$ and $\eta \pi^0$ modes) and $\Mbc>5.27$ GeV/$c^2$, respectively.
No significant signals are observed for the neutral $B$ meson modes; for these
modes we
set branching fraction upper limits at the 90\% confidence level. The  
upper limit for each mode is determined using the combined likelihood 
for the two $\eta$ decay channels with
the reconstruction efficiency reduced by $1 \sigma$. We vary
the signal PDF and the expected $\eta K^*$ feed-down in the $\eta K^0$ mode to 
compute the likelihood as a function of branching fraction; the largest 
branching fraction that covers 90\% of the 
likelihood area is chosen to be the upper limit.
 
\begin{figure}[htb]
\includegraphics[width=0.7\textwidth]{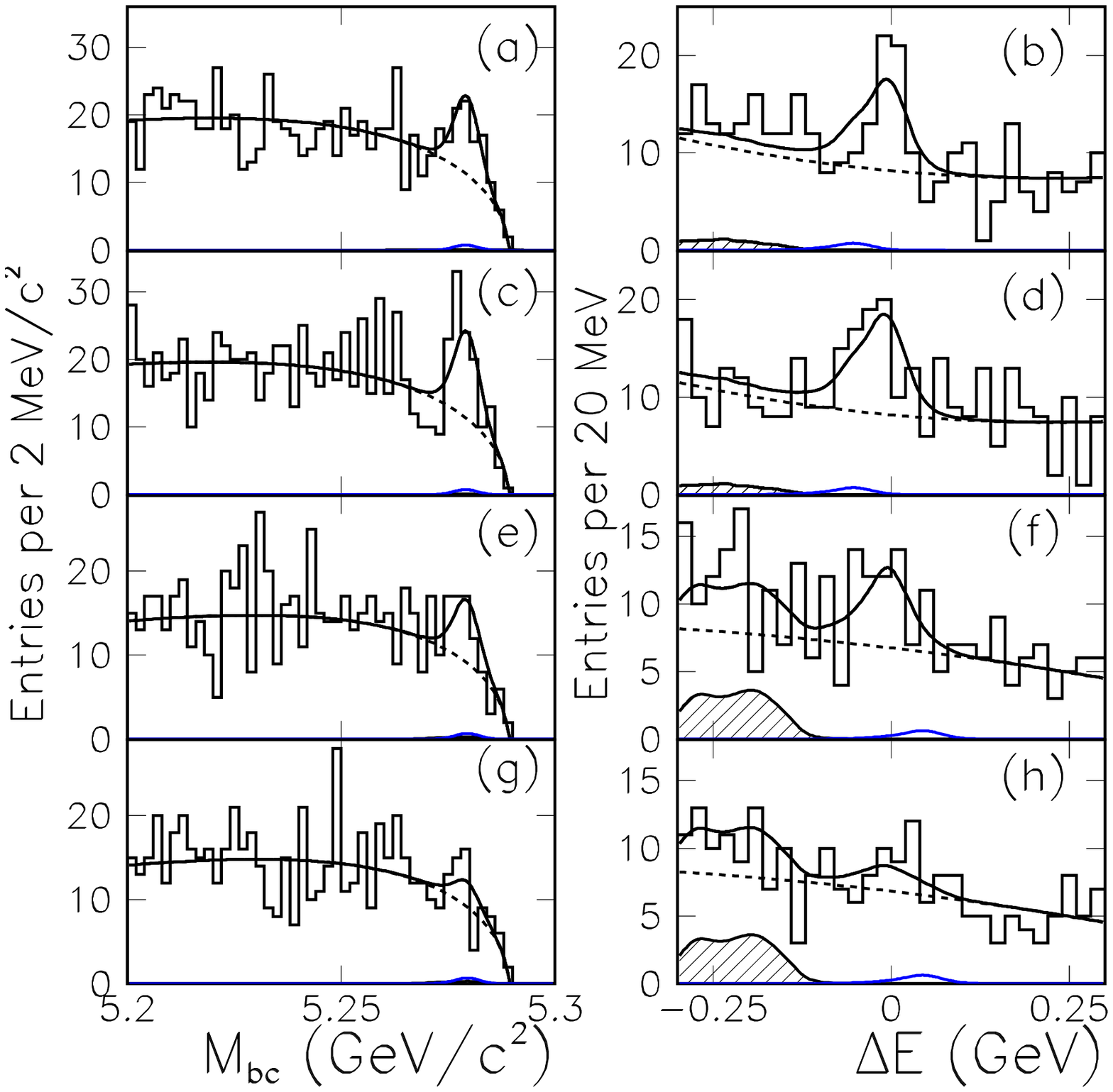}
\caption{$\Mbc$ and $\de$ projections for (a,b) $B^+\to\eta \pi^+$, (c,d) 
$B^- \to \eta\pi^-$,
(e,f) $B^+\to\eta K^+$, and (g,h) $B^-\to\eta K^-$  with 
the $\etagg$ and $\etapi$ modes combined. Open
histograms are data, solid curves are the fit functions, dashed lines show
the continuum contributions and shaded histograms are the $\eta K^*/\eta \rho$
contributions.  Small curves around $\Mbc = 5.28$ GeV/$c^2$ and $\de =\pm 0.05$
GeV are the reflection background on $\btoetapi$ and $\btoetak$.}  
\label{fig:acp}
\end{figure}

Significant signals are observed for charged $B$ decays. We investigate
their partial rate asymmetries by extracting signal yields separately from the 
 $B^+$ and $B^-$ samples. Unbinned maximum likelihood fits are performed 
independently for the two $\eta$ decay modes in order to reduce the 
systematic uncertainties. The same signal and background PDFs as used in the 
branching fraction measurement 
are applied. The parameters of the continuum PDF are fixed according to
the branching fraction results. Contributions from 
$\bb$ backgrounds are required to be equal for the $B^+$ and $B^-$ samples. 
 Figure \ref{fig:acp} shows the $\Mbc$ and $\de$ projections. 
  The $A_{CP}$ results for the two $\eta$ decay modes are combined
assuming that the errors are Gaussian. Systematic errors due to  
uncertainties  in the signal PDF are estimated by varying the peak positions and
resolutions. We also check the $A_{CP}$ values after varying the amount of the
expected $\eta K^*$ feed-down and the reflection background. The $\bb$ 
contributions are allowed to be different for the two samples to obtain
the systematic error. The largest 
uncertainty is the asymmetry of the reflection.  A possible detector bias in $A_{CP}$ 
is studied using $B\to D\pi^+$ decays. The obtained 
uncertainty is 0.5\%. Each $A_{CP}$ deviation is added quadratically to provide
the total systematic uncertainty.   
   
In summary, we have  observed  $\btoetapi$ and found evidence for 
$\btoetak$; the 
measured branching fractions and partial rate asymmetries are summarized
in Table \ref{tab:result}.  We conclude that the $B^+\to \eta \pi^+$ 
 branching fraction is larger than that of  $B^+\to \eta K^+$. 
The measured $\btoetapi$ branching fraction  is 
consistent with an earlier result published by the BaBar Collaboration; 
however, unlike the large negative $A_{CP}$
measured by BaBar, the central value in this analysis
is small and positive, and is consistent with no asymmetry. For the decay 
$\btoetak$, the measured 
branching fraction is 40\% lower than the published result of the BaBar 
experiment, corresponding 
to a 1.3 $\sigma$ deviation. It is interesting to note that although the 
errors are still large, both experiments  suggest
a large negative $A_{CP}$ value for $\btoetak$, which is anticipated by some
theories \cite{etakcp}. No significant signals are found in  neutral 
$B\to\eta h$ decays and upper limits at the 90\% confidence level are given.    
 
We thank the KEKB group for the excellent operation of the
accelerator, the KEK Cryogenics group for the efficient
operation of the solenoid, and the KEK computer group and
the National Institute of Informatics for valuable computing
and Super-SINET network support. We acknowledge support from
the Ministry of Education, Culture, Sports, Science, and
Technology of Japan and the Japan Society for the Promotion
of Science; the Australian Research Council and the
Australian Department of Education, Science and Training;
the National Science Foundation of China under contract
No.~10175071; the Department of Science and Technology of
India; the BK21 program of the Ministry of Education of
Korea and the CHEP SRC program of the Korea Science and
Engineering Foundation; the Polish State Committee for
Scientific Research under contract No.~2P03B 01324; the
Ministry of Science and Technology of the Russian
Federation; the Ministry of Education, Science and Sport of
the Republic of Slovenia; the National Science Council and
the Ministry of Education of Taiwan; and the U.S.\
Department of Energy.


%


\begin{thebibliography}{99}
\bibitem{CC}
Throughout this paper, 
the inclusion of the charge conjugate mode decay is implied
unless otherwise stated.
\end{thebibliography}
\end{document}